\documentclass[iop]{emulateapj}

\usepackage{multirow}
\usepackage{natbib}
\usepackage{epsfig}
\usepackage{graphics}
\usepackage{longtable,rotate}

\shorttitle{Broad Line Blazars from 1LAC}
\shortauthors{M. S. Shaw et al} 
		
\newcommand{\noprint}[1]{}
\newcommand{\figsetstart}{{\bf Fig. Set} }
\newcommand{\figsetend}{}
\newcommand{\figsetgrpstart}{}
\newcommand{\figsetgrpend}{}
\newcommand{\figsetnum}[1]{{\bf #1.}}
\newcommand{\figsettitle}[1]{ {\bf #1} }
\newcommand{\figsetgrpnum}[1]{\noprint{#1}}
\newcommand{\figsetgrptitle}[1]{\noprint{#1}}
\newcommand{\figsetplot}[1]{\noprint{#1}}
\newcommand{\figsetgrpnote}[1]{\noprint{#1}}

\slugcomment{Accepted for publication in ApJ}

\usepackage{lscape}

\begin{document}

\title{Spectroscopy of Broad Line Blazars from 1LAC}
\setlength{\paperheight}{11.0in}
\setlength{\paperwidth}{8.5in}

\author{Michael S. Shaw\altaffilmark{1}, Roger W. Romani\altaffilmark{1}, Garret Cotter\altaffilmark{2}, Stephen E.\ Healey\altaffilmark{1}, 
Peter F.\ Michelson\altaffilmark{1}, 
Anthony C.\ S.\ Readhead\altaffilmark{3}, Joseph L.\ Richards\altaffilmark{3}, 
Walter Max-Moerbeck\altaffilmark{3}, Oliver G.\ King\altaffilmark{3}, William J.\ Potter\altaffilmark{2}
}

\altaffiltext{1}{Department of Physics/KIPAC, Stanford University, Stanford, CA 94305}
\altaffiltext{2}{Department of Astrophysics, University of Oxford, Oxford OX1 3RH, UK}
\altaffiltext{3}{Department of Astronomy, California Institute of Technology, Pasadena, CA 91125}

\begin{abstract}
We report on optical spectroscopy of 165 Flat Spectrum Radio Quasars (FSRQs) 
in the {\it Fermi} 1LAC sample, which have helped allow a nearly complete study
of this population.  {\it Fermi} FSRQ show significant evidence for non-thermal
emission even in the optical; the degree depends on the $\gamma$-ray hardness.
They also have smaller virial estimates of hole mass than the optical quasar 
sample. This appears to be largely due to a preferred (axial) view of the 
$\gamma$-ray FSRQ and non-isotropic ($H/R \sim 0.4$) distribution of broad-line
velocities. Even after correction for this bias, the {\it Fermi} FSRQ show 
higher mean Eddington ratios than the optical population. A comparison of 
optical spectral properties with Owens Valley Radio Observatory radio flare 
activity shows no strong correlation.
\end{abstract}


\section{Introduction}

The {\it Fermi} Gamma-Ray Space Telescope was launched on 2008 June 11. Its 
primary instrument is the Large Area Telescope \citep[LAT]{atw09}. {\it Fermi} 
generally operates in sky survey mode, observing the entire sky every $3$ hours,
and providing approximately uniform sky coverage on time scales of days to 
years.

The {\it Fermi} LAT First Source Catalog \citep[1FGL]{1FGL} catalogs the 1451 
most significant sources detected in {\it Fermi}'s first year of operation. 
Based on the 1FGL catalog, The First Catalog of AGN Detected by the {\it Fermi}
LAT \citep[1LAC]{1LAC} is the largest radio-$\gamma$ selected sample of blazars
to date, associating 671 $\gamma$-ray sources to $709$ AGN (some may be 
unresolved composites) in the high-latitude sample.
 
Our quest is to optically characterize these sources, seeking maximum 
completeness in spectroscopic identifications and using the spectra to 
constrain the properties of these AGN. Optically, the {\it Fermi} sources are 
evenly split between Flat Spectrum Radio Quasars (FSRQs) and BL Lacerate Objects
(BL Lacs). In this paper, we focus on the FSRQs. A companion paper (Shaw et al,
in prep.) addresses the BL Lac objects.

In \S \ref{sec:obs}, we discuss the observational program and the data reduction
pipeline. In \S \ref{sec:analysis}, we describe the measurements and derived 
data products. In \S \ref{sec:discussion}, we measure the continuum emission and
non-thermal pollution. In \S \ref{sec:masses}, we estimate the black hole masses and Eddington ratio
of the {\it Fermi} FSRQ. In \S \ref{sec:orientation}, we discuss the orientation and shape of the
the broad line regions in this population, and in \S \ref{sec:multiwavelength},
we relate this data set to on-going radio monitoring of these AGN.

In this paper, we assume an approximate concordance cosmology--- 
$\Omega_m=0.3$, $\Omega_\Lambda=0.7$, and $H_0=70\ $km s$^{-1}$ Mpc$^{-1}$.

\section{Observations and Data Reduction}
\label{sec:obs}

\subsection{The FSRQ Sample}
\label{sec:sample}

This paper reports on a multi-year observing campaign to follow-up the {\it 
Fermi} blazars. A principal aim is to achieve high redshift completeness 
for the 1LAC sample \citep{1LAC}. 

In this paper, we discuss the spectra of FSRQs and other LAT blazar associations
with strong emission lines.  A major contribution is new spectroscopy of 165 
of these blazars. To extend the analysis, we also measured archival spectra of 
64 SDSS blazars in the sample, for a total of 229 spectra.

This work takes the 1LAC high latitude sample to 96\% type completeness, with 
316 FSRQs, 322 BL Lacs, 33 other AGNs, 4 LINERs, and 4 Galaxies. There are 30 
remaining associated flat spectrum radio sources of unknown type -- generally 
these represent objects that are optically extremely faint (R $>$ 23) or show faint continuum-dominated spectra, where current spectroscopy does not 
have sufficient S/N to unambiguously confirm a BL Lac-type ID.

The most important sub-set of this emission-line sample are the objects with
traditional FSRQ properties -- in addition to the flat spectrum radio core
emission which allows the LAT counterpart association, we require emission 
lines with kinematic FWHM $> 1000 $ km s$^{-1}$ and bolometric luminosity 
$> 10^{42}$ erg s$^{-1}$. We find that 188 FSRQ meet these criteria, including
10 low-latitude sources with 1LAC FSRQ associations. In addition, some 11 BL 
Lacs show well-detected broad lines. For this paper we adopt the traditional 
heuristic BL Lac definition: continuum-dominated objects with observed frame 
line equivalent width (EW) of $<5$ \AA\ and, where measured, Balmer break strength of $<\ $0.5 \citep{cgrabs}.
We classify an object as a `BL Lac' if it meets these spectroscopic
criteria at any epoch. For 6 of the 11 BL Lac our spectra includes epochs in a
`low' state where decreased continuum reveals broad emission lines with
$>5$ \AA\ EW. The other 5 objects satisfy the BL Lac criteria in all of
our spectra, but nevertheless show highly significant, albeit low EW, broad
lines.

The emission line sample contains 29 other objects -- spectroscopically these are 
9 galaxies, 5 LINERs, and 15 other AGN. These show only strong narrow lines. 
While the line strengths rule out BL Lac IDs for these objects, they are 
manifestly different from our typical FSRQs. They may be misaligned radio 
galaxies, intrinsically weak AGN, narrow-line Seyferts and other less common
types of $\gamma$-ray emitters \citep{misaligned}.

Note that this sample of digital spectra does not include measurements for a 
number of bright, famous blazars, with historical spectroscopic classifications
in the literature. Indeed, 137 of the 316 {\it Fermi} FSRQs fall into this 
category. Measured in $R$, the sources with archival spectra are brighter by 
1.4 magnitudes than our sample, so their omission may introduce systematic 
effects, discussed briefly in Section \ref{sec:masscomp}.

\subsection{Observations}

We have used medium and large telescopes in both hemispheres in a many-faceted 
assault on this sample. Observations were obtained from the Marcario Low 
Resolution Spectrograph (LRS) on the Hobby-Eberley Telescope (HET), the Large 
Cass Spectrometer (LCS) on the 2.7m at McDonald Observatory, on the ESO Faint 
Object Spectrograph and Camera \citep[EFOSC2]{buz84} and ESO Multi-Mode 
Instrument \citep[EMMI]{dek86} at the New Technology Telescope at La Silla 
Observatory (NTT), on the Double Spectrograph (DBSP) on the 200'' Hale 
Telescope at Mt. Palomar, on the FOcal Reducer and low dispersion Spectrograph 
\citep[FORS2]{app98} on the Very Large Telescope at Paranal Observatory (VLT),
and on the Low Resolution Imaging Spectrograph (LRIS) at the W. M. Keck 
Observatory (WMKO). Observational configurations and objects observed are 
listed in Table \ref{table:obs}.

The observing runs jointly targeted emission line objects discussed in this 
paper and BL Lacs, to be discussed in a companion paper (Shaw et al, in prep.).
All objects discussed in this work have highly significant emission lines and 
confirmed spectroscopic redshifts. 

All spectra are taken at the parallactic angle, except for LRIS spectra using 
the atmospheric dispersion corrector, where we observed in a north-south 
configuration. In a few cases, we rotated the slit angle to minimize 
contamination from a nearby star. At least two exposures are taken of every 
target for cosmic ray cleaning. Typical exposure times are $2$x$600$ s.

With the variety of telescope configurations and varying observing conditions, 
the quality of the spectra are not uniform: resolutions vary from $4$ to $15$ 
\AA, exposure times from $360$ s to $2400$ s, and telescope diameters from 
$2.7$ m to $10$ m. During this campaign, we generally used the minimum exposure 
required for a reliable redshift, rather than exposure to a uniform S/N. This 
may introduce selection effects into the sample, discussed briefly in 
\S \ref{sec:discussion}.

\begin{deluxetable*}{cccccccc}
\tabletypesize{\small}
\tablecaption{Observing Configurations}
\tablehead{
\colhead{Telescope} & \colhead{Instrument} & \colhead{Resolution} & \colhead{Slit Width} & \colhead{Objects} & \colhead{Filter} & \colhead{$\lambda_{min}$} & \colhead{$\lambda_{max}$} \\
\colhead{} & \colhead{} & \colhead{\AA} & \colhead{Arcseconds} & \colhead{} & \colhead{} & \colhead{\AA} & \colhead{\AA}
}
\startdata
HET & LRS & 15 & 2 & 77 & GG385 & 4150 & 10500\\
HET & LRS & 8 & 1 & 1 & GG385 & 4150 & 10500 \\
McD 2.7m & LCS & 15 & 2 & 1 & - & 4200 & 8200 \\
NTT & EFOSC2 & 16 & 1 & 14 & - & 3400 & 7400 \\
NTT & EMMI & 12 & 1 & 8 & - & 4000 & 9300 \\
Palomar 200'' & DBSP & 5 / 15 & 1 & 3 & - & 3100 & 8100\\
Palomar 200'' & DBSP & 5 / 15 & 1.5 & 1 & - & 3100 & 8100\\
Palomar 200'' & DBSP & 5 / 9 & 1.5 & 2 & - & 3100 & 8100\\
VLT & FORS2 & 11 & 1 & 22 & - & 3400 & 9600 \\
VLT & FORS2 & 17 & 1.6 & 2 & - & 3400 & 9600\\
WMKO & LRIS & 4 / 7 & 1 & 11 & - & 3100 & 10500 \\
WMKO & LRIS & 4 / 9 & 1 & 22 & - & 3100 & 10500
\enddata
\tablecomments{For DBSP and LRIS the blue and red channels are split by a 
dichroic at 5600 \AA; the listed resolutions are for blue and red side,
respectively.}
\label{table:obs}
\end{deluxetable*}

\subsection{Data Reduction Pipeline}

Data reduction was performed with the IRAF package \citep{tod86, val86} using 
standard techniques. Data was overscan subtracted, and bias 
subtracted. Dome flats were taken at the beginning of each night, the spectral 
response was removed, and all data frames were flat-fielded.  Wavelength 
calibration employed arc lamp spectra and was confirmed with checks of night sky
lines. For these relatively faint 
objects, we employed an optimal extraction algorithm \citep{val92} to maximize 
the final signal to noise. For HET spectra, care was taken to use sky windows 
very near the longslit target position so as to minimize spectroscopic 
residuals caused by fringing in the red, whose removal is precluded by the
rapidly varying HET pupil. Spectra were visually cleaned of residual cosmic 
ray contamination affecting only individual exposures.  

We performed spectrophotometric calibration using standard stars from 
\citet{oke90} and \citet{boh07}. In most cases standard exposures were available
from the data night.  Since the HET is queue scheduled, standards from 
subsequent nights were sometimes used. At all other telescopes, multiple 
standard stars were observed per night under varying atmospheric conditions and
different air-masses. The sensitivity function was interpolated between standard 
star observations when the solution was found to vary significantly with time.

For blue objects, broad-coverage spectrographs can suffer significant second 
order contamination. In particular, the standard HET configuration using a
Schott GG385 long-pass filter permitted second-order effects redward of 7700 \AA.
The effect on object spectra were small, but for blue WD spectrophotometric
standards, second order corrections were needed for accurate determination of
the sensitivity function. This correction term was constructed following
\citet{fors}. In addition, since BL Lac spectra are generally simple power laws,
we used BL Lacs observed during these runs (Shaw et al, in prep.) to monitor
second order contamination and residual errors in the sensitivity function.
This resulted in excellent, stable response functions for the major data sets.

Spectra were corrected for atmospheric extinction using standard values. We 
followed \citet{kri87} for WMKO LRIS spectra, and used the mean KPNO extinction 
table from IRAF for P200 DBSP spectra. Our NTT, VLT, and HET spectra do not
extend into the UV and so suffer only minor atmospheric extinction. These
spectra were also corrected using the KPNO extinction tables.
We removed Galactic extinction using IRAF's de-reddening function and the Schlegel 
maps \citep{sch98}. We made no attempt to remove intrinsic reddening (ie. from 
the host galaxy).

Telluric templates were generated from the standard star observations in each 
night, with separate templates for the oxygen and water line complexes. We 
corrected separately for the telluric absorptions of these two species. We found
that most telluric features divided out well, with significant residuals only
apparent in spectra with very high S/N. On the HET spectra, residual
second order contamination prevented complete removal of the strong water band 
red-ward of $9000$ \AA.

	When we had multiple epochs of these final cleaned, flux-calibrated 
spectra with the same instrumental configuration, we checked for strong
continuum variation. Spectra with comparable fluxes were then combined
into a single best spectrum, with individual epochs weighted by S/N.

Due to variable slit losses and changing conditions between object and 
standard star exposures, we estimated that the accuracy of our absolute 
spectrophotometry is $\sim 30\%$ \citep{cgrabs}, although the relative 
spectrophotometry is considerably better.

Redshifts were generally confirmed by multiple emission lines and derived by 
cross-correlation analysis using the rvsao package \citep{rvsao}. For a few 
objects only single emission lines were measured with high S/N. In general
we could use the lack of otherwise expected features to identify the species
and the redshift with high confidence. Nevertheless, single-line redshifts are 
marked (Tables \ref{tab:lowz} and \ref{tab:highz}) with a colon (:) and discussed in \S \ref{sec:obs_individual}.
Velocities are not corrected to helio-centric or LSR frames.

\begin{figure*}
\epsscale{1.4}
\vspace{-30pt}
\hspace*{-40pt}
\plotone{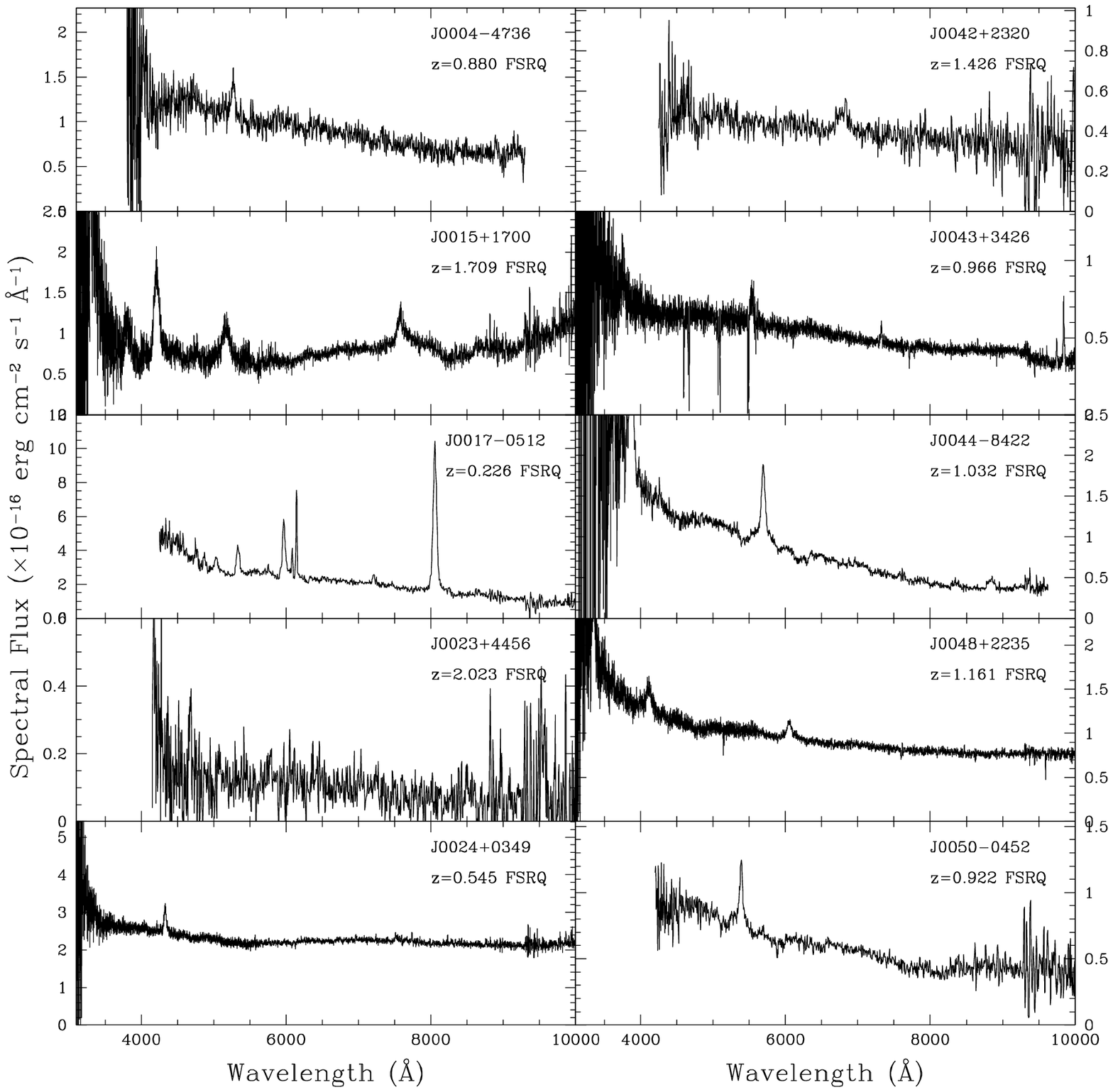}
\vspace{-50pt}
\caption{Spectra of the {\it Fermi} blazars reported on in this work, ordered by RA. Redshifts and optical types are listed for each spectrum. All spectra available in electronic figure set.}
\label{fig:spectra}
\end{figure*}

Reduced spectra of the newly measured objects are presented in Figure \ref{fig:spectra} (full figure
available on-line).

\figsetstart
\figsetnum{1}
\figsettitle{Spectra}

\figsetgrpstart
\figsetgrpnum{1.1}
\figsetgrptitle{J0004-4736 to J0050-0452}
\figsetplot{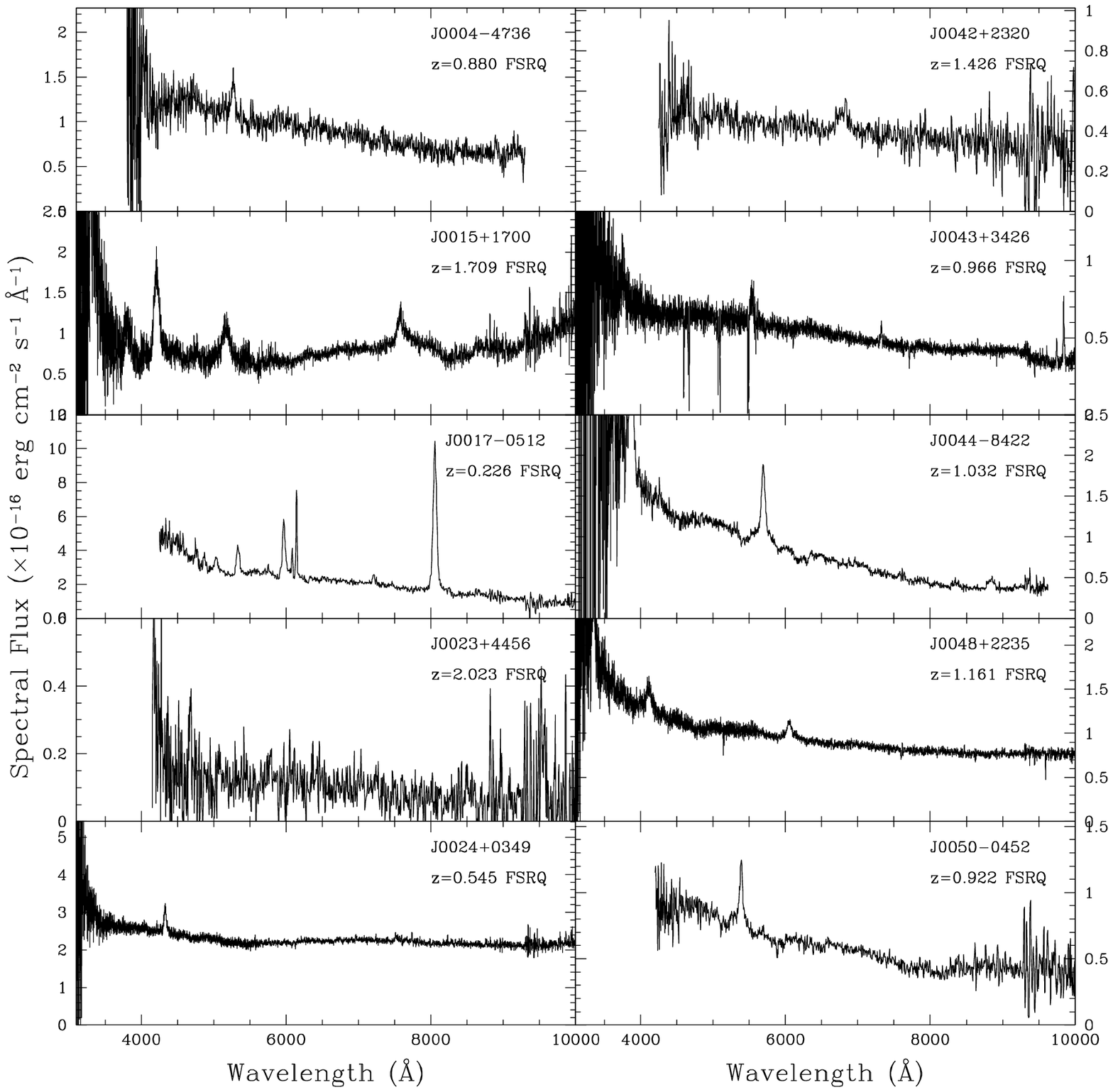}
\figsetgrpnote{Spectra of the {\it Fermi} blazars reported on in this work, ordered by RA. Redshifts and optical types are listed for each spectrum.}
\figsetgrpend

\figsetend

\subsection{Individual Objects}
\label{sec:obs_individual}

A few observations require individual comment:

J0023+4456: This spectrum had low S/N ($\sim 2$) and, while \ion{C}{4}, 
\ion{C}{3}, and \ion{Mg}{2} lines were tentatively identified, we mark
the final solution as uncertain.

J0654+5042: This blazar, $0.5$'' from a bright star, was observed using DBSP 
with the slit along the blazar-star axis. We were able to isolate the stellar
light by extraction from the wings of the convolved object. This spectrum
was scaled and subtracted from the blazar spectrum to remove the stellar rest
wavelength absorption features, giving a relatively clean blazar spectrum. While
the emission line and redshift measurements are unambiguous, the continuum
spectrophotometry should be treated with caution.

J0949+1752, J1001+2911, J1043+2408, and J1058+0133: These blazars show only a
single broad emission line; good spectral coverage gives high confidence of
identification as \ion{Mg}{2}.

J1330+5202: This blazar shows only strong [\ion{O}{2}] in emission at $3727$ \AA,
with associated host Ca H/K absorptions. No broad emission lines are observed.

J1357+7643: Only one broad line is detected with high significance. A weaker
corroborating feature implies \ion{C}{4}, \ion{Mg}{2} at z=1.585. However,
a z=0.431 solution cannot be ruled out so we mark this solution as tentative.

J2250-2806: With a single strong narrow line (assigned to [\ion{O}{2}] $3727$ \AA),
this blazar has an uncertain redshift. No broad emission lines are observed.

\section{Measured Properties}
\label{sec:analysis}

\subsection{Primary Spectra and Multiple Epochs}
\label{sec:mult}

For some objects, multiple exposures at different epochs show significant
spectral variability. In general, follow-up epochs have higher resolution or
S/N, or targeted the blazar in a lower flux state. In a few cases SDSS spectra 
were published after we had obtained independent data. We adopt the spectrum 
showing the highest S/N for broad line detection as the primary spectrum used
in this analysis. In general, this is the spectrum used to solve for the source
redshift.  As noted above, for particularly faint objects we can make 
S/N-weighted combinations of spectra (for which no strong flux variation is
seen) to obtain the best primary spectrum. This is particularly valuable for
the HET data, as limited track time requires short observations, but queue
scheduling with very stable configurations allows easy combination of multiple
epochs.

	As noted above, in six cases, sources in this sample
transitioned from a bright (continuum dominated) state to a lower state.
For these the `Primary' spectrum is from the low state, allowing better line 
measurements, during which the source is a nominal FSRQ (broad line EW
$>$5 \AA). We nevertheless retain the original BL Lac typing (see
\S \ref{sec:individual}). Clearly a more physical separation between 
the classes, and attention to duty cycles in the various `states' is needed.

\subsection{Continuum Properties}
\label{sec:contprop}

We have fit power law continua to all our spectra, using the 
scipy.optimize.leastsq \citep{scipy} routine based on Levenberg-Marquardt 
fitting to estimate parameter values and statistical errors. For these fits
we first excise regions around the strongest emission lines in the blazar rest 
frame: $\lambda\lambda$1200--1270, 1380--1420, 1520--1580, 1830--1970, 
2700--2900, 3700--3750, 4070--4130, 4300--4380, 4800--5050, 6500--6800 \AA.
We then fit simple power law flux spectra to the entire spectral range,
excluding regions at the blue or red end with uncharacteristically large noise.
We do not here attempt to separate the thermal disk contribution. The results
are presented as power law indices, $\alpha$, and continuum fluxes at 
$10^{14.7}$\,Hz ($\sim 5980$ \AA), the center of our spectral range, as measured in the observer frame. These data may be combined with multi wavelength data to study the SED of the blazers in our sample.

The statistical errors on $F_\nu$ (Tables \ref{tab:lowz} and \ref{tab:highz}) are generally small, but should be
convolved with the overall fluxing errors, estimated at $30\%$. We estimate 
errors on the spectral index by independently fitting the red and blue halves 
of the measured spectrum. We then sum the differences in quadrature for an
estimated error:
\begin{equation}
\alpha_\textrm{err} = \left((\alpha_\textrm{red} - \alpha)^2 + (\alpha_\textrm{blue} - \alpha)^2\right)^{1/2}
\end{equation}
Note that large errors bars generally indicate deviation from power-law 
continua rather than poor statistics. We find that extreme values of $\alpha$ 
seem to correlate with large Galactic $A_V$, suggesting errors in the assumed 
extinction and residual curvature in the corrected spectrum.

\subsection{Line Properties}

All line measurements are conducted in the object's rest frame. The local
continuum and line parameters are measured with the scipy.optimize.leastsq 
routines. We first isolate the local continua surrounding the line, as in \S 
\ref{sec:contprop} (including an additional pseudo-continuum for \ion{Mg}{2} 
from the \ion{Fe}{2}/\ion{Fe}{3} line complex). Then, the line is fit to the 
continuum-subtracted data.  Reported errors are purely statistical.

We follow \citet{she11} in their line measurement techniques to facilitate 
comparison of our results with with measurements in the extensive SDSS quasar 
catalog. We focus on lines used for black hole mass estimates: H$\beta$ at 
$4861$ \AA, \ion{Mg}{2} at $2800$ \AA, and \ion{C}{4} at $1550$ \AA. 

Before reporting kinematic widths, we subtract in quadrature the observational 
resolution, to eliminate bias between objects measured by different telescopes,
and to improve estimates of narrower lines. In higher redshift sources, emission
lines are sometimes contaminated by intervening or associated absorption line 
systems. We visually check all emission lines.

We found strong absorption lines (associated and self-absorption systems) in a number of objects and edited these out before fitting the emission lines. Associated \ion{Mg}{2} absorbers were found in J0043+3426 ($\Delta v \sim - 2400$ km/s), J0252-2219 ($\Delta v \sim -3500$ km/s and $\Delta v \sim -2800$ km/s), J1120+0704 ($\Delta v \sim -1600$ km/s), J1639+4705 ($\Delta v \sim - 600$ km/s) and J2212+2355 ($\Delta v \sim - 2400$ km/s). In J0325+2224 we find self-absorbed \ion{C}{4} absorption. In J2321+3204, we find two systems ($\Delta v \sim - 400$ km/s and $\Delta v \sim 100$ km/s) and remove the absorptions from both \ion{Mg}{2} and \ion{C}{4} before fitting.  Finally, in J2139-6732, we edit out an intervening ($z = 0.923$) \ion{Fe}{2} absorption system coincident with the \ion{C}{4} emission line. We next summarize particular issues for the three major species fit.

\subsection{Fitting H$\beta$}
\label{sec:hb}

For H$\beta$, it is essential to include narrow components in the line fit. 
After power law continuum removal, we simultaneously fit broad and narrow 
H$\beta$, narrow [\ion{O}{3}]4959 \AA, and narrow [\ion{O}{3}]5007 \AA
\citep{mcl04}.  We fix the rest wavelengths of narrow H$\beta$ and the 
[\ion{O}{3}] lines at the laboratory values and fix widths at the spectral
resolution, as measured from sky lines; the broad H$\beta$ center and width are
free to vary. All lines are modeled with single Gaussian profiles. The 
continuum is measured at 5100 \AA.

\subsection{Fitting \ion{Mg}{2}}
\label{sec:MgII}

For \ion{Mg}{2}, we fit the pseudo-continuum from broad \ion{Fe}{2}/\ion{Fe}{3}
along with a power law. For the former, we use a template \citep{ves01} 
convolved with the observational resolution. The power law is fit using the
$\lambda\lambda$2200--2700 and 2900--3100 \AA\, portions of the spectrum; both
continuum component fluxes and the power law index vary freely.

After continuum subtraction, we fit the \ion{Mg}{2} line with broad and
narrow Gaussian components. For this line, the continuum measurement is made 
at 3000 \AA, to minimize residual Fe contamination \citep{mcl04}. 

\subsection{Fitting \ion{C}{4}}
\label{sec:civ}

Here, we fit the power law continuum over $\lambda\lambda$1445--1464 \AA\, and 
1700--1705 \AA. After subtraction, we fit the \ion{C}{4} line with three 
Gaussians, following \citet{she11}, and report the full FWHM of the line. 
Any narrow self-absorption components are visually identified and removed
prior to the fit, as with the intervening absorbers described above.

\subsection{Objects with Special Type Classifications}
\label{sec:individual}

Several blazars were classified as BL Lacs in initial epoch observations. At 
the `primary' spectrum epoch, with low continuum, each was a nominal FSRQ. The
objects which changed (and continuum decrease) were: J0058+3311 ($8\times$),
J0923+4125 ($4\times$), J1001+2911 ($6\times$), J1607+1551 ($5\times$),
J2031+1219 ($4\times$) and J2244+4057 ($10\times$).

	With very high S/N observations, we were able to detect broad lines
at high significance at EW levels $<5$ \AA\ in several objects. These were
thus `BL Lacs' at all of our epochs, but can be analyzed along with the FSRQ.
The BL Lacs (and strongest broad line EWs) were:
J0430-2507 (\ion{Mg}{2} at EW=0.9 \AA), J0516-6207 (\ion{C}{4} at EW=1.6 \AA; 
\ion{C}{3}, \ion{Mg}{2} also present), J1058+0133 (\ion{Mg}{2} at EW=2.2 \AA),
J2236+2828 (\ion{Mg}{2} at EW=4.9 \AA) and J2315-5018 (\ion{Mg}{2} at EW=3.8 \AA).
These EW measurements are in observed frame. Clearly these sources are 
transitional between our standard FSRQ and BL Lac types.

\subsection{Calibrating our Measurements}

For 53 of the 64 SDSS spectra in our sample \citet{she11} tabulate standard
measurements of the continuum and emission lines properties. This allows us to
check the consistency of our measurements techniques.

We find that our measurements of continuum and line luminosities are consistent
with those of \citet{she11}, although some scattered differences in line luminosity
are seen, as the luminosity is sensitive to the fitting details. In particular,
for \ion{Mg}{2} and H$\beta$, the variations in the narrow line component flux 
can change the measured broad line luminosity. Nevertheless, we see no 
systematic offset between our results and those of \citet{she11}. 

However, we do note that our fitted kinematic FWHM values are systematically
$\sim 10$\%  larger (for the same sources) than those of \citet{she11}.
For \ion{Mg}{2} and H$\beta$, we attribute this to small differences in the fit
to the narrow component. For \ion{C}{4}, three Gaussians are used and we
suspect that our central (narrowest) Gaussian is systematically weaker than that 
found in the SDSS fits. As we shall see, interesting differences from the SDSS quasars include
narrower lines, on average, so we conservatively chose not to bias our
fitter to force agreement. Nevertheless FWHM effects may be suspect at the 
10\% scale.

\subsection{Comparison to SDSS Non-Blazars}
\label{sec:sdsscomp}

\begin{figure*}
\epsscale{1.4}
\vspace{-30pt}
\hspace*{-50pt}
\plotone{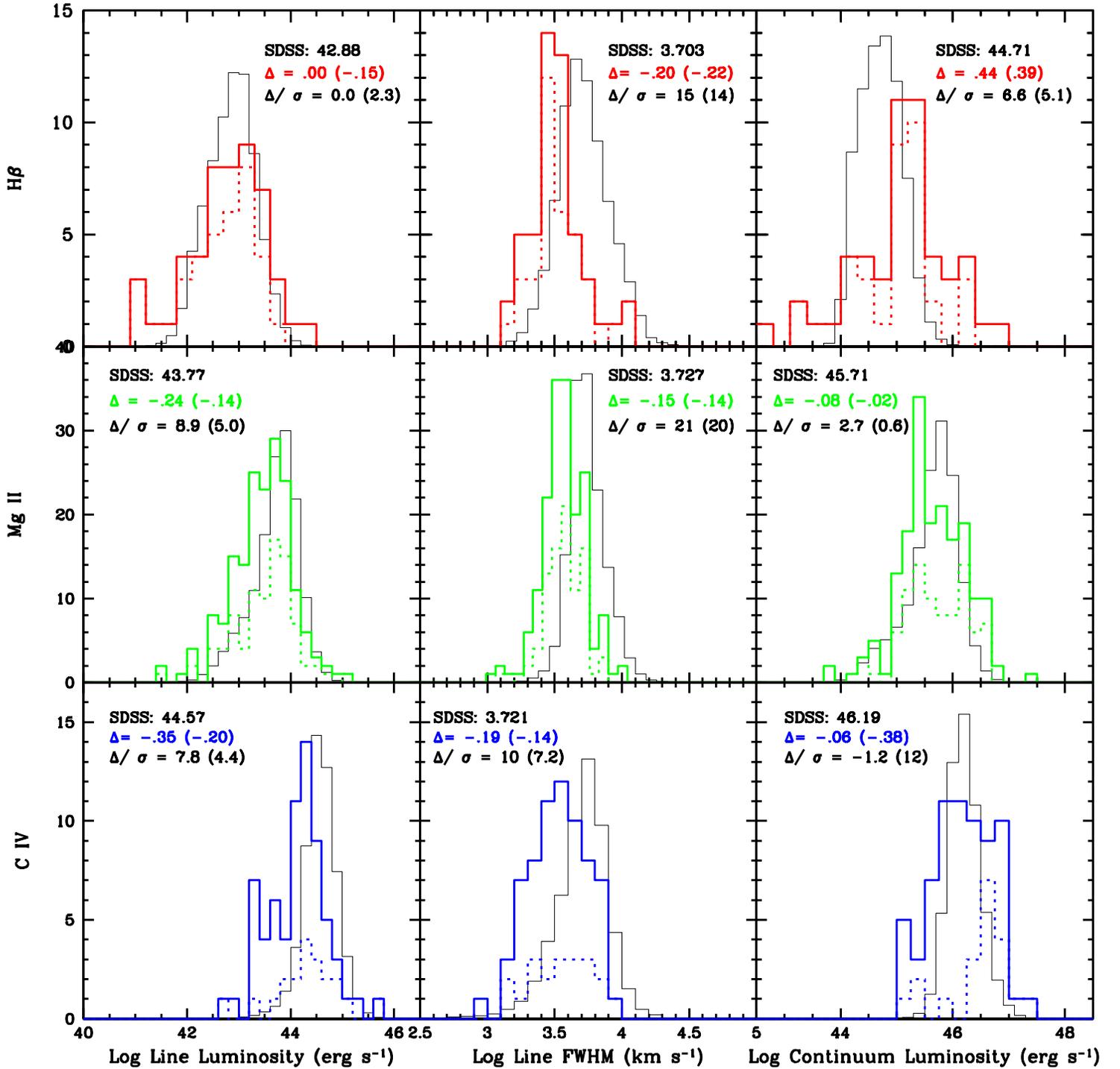}
\vspace{-30pt}
\caption{Line and continuum properties of our 1LAC sample (thick colored lines) 
compared to the SDSS clean sample \citep{she11} (thin black lines). To match the
flux limit of the clean sample, we additionally plot the subset of {\it Fermi}
objects  with magnitude $< 19.1$ (dotted lines). For each species and parameter,
the plots are labeled with the SDSS mean value and the offset $\Delta$ and
estimated statistical significance $\Delta/\sigma$ of the means of the 
{\it Fermi} FSRQ, as measured in log space. The corresponding properties for
the flux-limited sub-sample are given in parentheses.}
\label{fig:raw}
\end{figure*}

To place our {\it Fermi} sample in the context of the broader QSO population,
we compare our measurements to those of the largely ($> 90\%$) radio quiet SDSS DR7 quasar
sample \citep{she11}. We find significant offsets in the mean values of
the continuum luminosities, line luminosities, and line FWHMs of the two classes
as plotted in Figure \ref{fig:raw}. Since our sample extends fainter than the 
SDSS QSOs, we also plot the distributions for a magnitude cut $i < 19.1$ sample
 \citep{sch10}.

To quantify these differences, we compute the medians, as well as the semi 
interquartile range (SIQR) for each distribution. We can use these values to
estimate statistical errors on the median as $\sigma = SIQR / \sqrt{n}$, where 
$n$ is the number of points in the sample. In Figure \ref{fig:raw}, we report 
both the median offset, $\Delta$, and its estimated significance, $\Delta / 
\sigma$, for each parameter. The corresponding medians and errors for the 
magnitude-cut sample are given in parentheses.

Several distributions show large offsets from the SDSS population. Our 
\ion{C}{4} and \ion{Mg}{2} lines are significantly less luminous than those
of the SDSS QSOs. Comparing the values for the flux-cut sample, we see that
a significant difference persists. We conclude that we are sampling a population
with intrinsically weaker lines or that fainter, weaker-lines objects are
lifted into our samples by the addition of extra  non-thermal continuum flux.
It seems that the local continua around \ion{Mg}{2} and \ion{C}{4} are
similar for our objects; we do notice that our
H$\beta$ continua seem substantially brighter than those of the QSO population.
We suspect that a lower redshift distribution, and larger host-light
contribution (see below) may explain this effect.

For all species (and for the sub-sample) we find that our measured lines are 
significantly narrower than those the SDSS average. We attribute this to 
preferential alignment, further discussed in \S \ref{sec:orientation}.

\section{Excess Continuum}
\label{sec:discussion}

We expect our radio-$\gamma$ selected population to have significant continuum 
contribution from non-thermal jet emission. At low redshift, there may also be 
additional galaxy light, especially around the H$\beta$ line, to the red of 
the Balmer break.

\subsection{Estimated Continua from Lines}
\label{sec:contfromlines}

In order to characterize the non-thermal emission in our spectra, we estimate 
a predicted continuum luminosity from the emission line fluxes, scaling to a 
non-blazar sample. \citet{she11} reports line and continuum luminosities for 
the DR7 quasars. These are highly correlated (rxy = 0.85, 0.91, and 0.68 for 
H$\beta$, \ion{Mg}{2}, and \ion{C}{4} respectively), and they fit continuum 
luminosities to line luminosities for \ion{Mg}{2} and \ion{C}{4}, finding 
\begin{equation}
\label{eqn:mglum}
\log L_{3000} = 1.016 \pm 0.003 \cdot \log L_{\textrm{\ion{Mg}{2}}} + 1.22 \pm 0.11
\end{equation}
\begin{equation}
\label{eqn:civlum}
\log L_{1350} = 0.863 \pm 0.009 \cdot \log L_{\textrm{\ion{C}{4}}} + 7.66 \pm 0.41
\end{equation}
We extend this treatment by fitting the SDSS QSO results for the H$\beta$ line:
\begin{equation}
\label{eqn:hblum}
\log L_{5100} = 0.802 \pm 0.049 \cdot \log L_{\textrm{H}\beta} + 1.574 \pm 0.060
\end{equation}

Since the line flux continuum predictions are calibrated to a non-blazar sample,
they should be independent of the non-thermal jet activity. Of course for both
samples, at low redshift, there may still be galaxy light contaminating the 
H$\beta$ region, increasing the scatter. We compare the predicted to observed continuum levels for
our {\it Fermi} FSRQ, finding that the lines predict (on average) $44\%$, 
$40\%$, and $32\%$ of observed continuum luminosities for H$\beta$, \ion{Mg}{2},
and \ion{C}{4} respectively. We attribute the excess to non-thermal `jet'
emission.

It is of interest to inspect the non-thermal dominance ($NTD\equiv 
L_{obs}/L_{pred}$) for individual objects in our sample. While most objects have
$NTD \sim 1$ (i.e. mostly thermal continua), for 57 blazars we have $NTD>2$
and 18 spectra show $NTD>10$ based on measurements of at least one line. We find 
significant correlation in same-object $NTD$ measured from the three species 
(rxy = 0.51 for \ion{Mg}{2} and \ion{C}{4}; rxy=0.73 for H$\beta$ and 
\ion{Mg}{2}) H$\beta$ generally measures larger $NTD$, which we attribute to 
contaminating galaxy light.   Thus for $10\%$ of our sample, the spectra are 
BL Lac-like and analysis of the optical properties must account for this 
dominant non-thermal emission.

\begin{figure}
\vspace{-0pt}
\hspace*{-20pt}
\epsscale{1.3}
\plotone{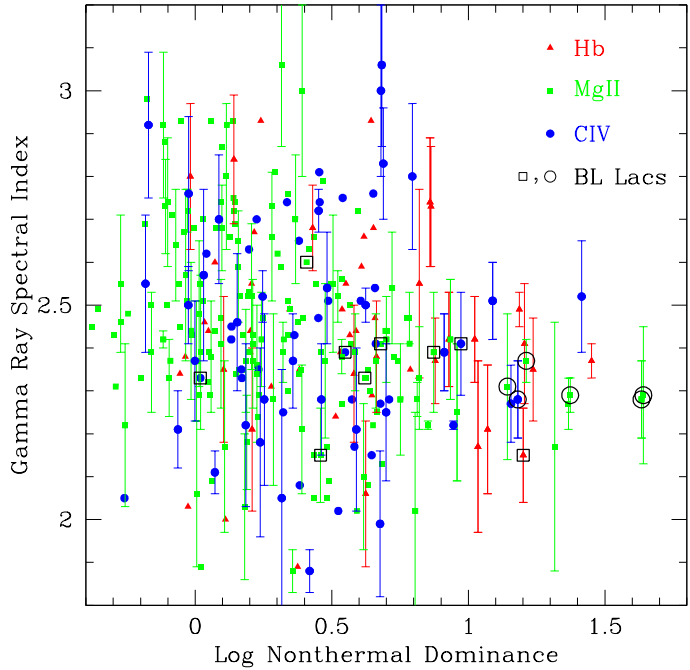}
\vspace{-0pt}
\caption{The non-thermal dominance is the ratio of the measured to the predicted continuum emission near each of the lines. The H$\beta$ sample may be contaminated by galaxy luminosity as well. BL Lacs in the primary spectrum are circles. Squares indicate BL Lac designations from a previous spectrum. Large non-thermal dominance is weakly correlated to low spectral indices in {\it Fermi} observations. We thin error bars for $NTD < 0.8$ for readability.} 
\label{fig:nontherm}
\end{figure}

\begin{figure*}
\vspace{-20pt}
\hspace*{-25pt}
\epsscale{1.25}
\plotone{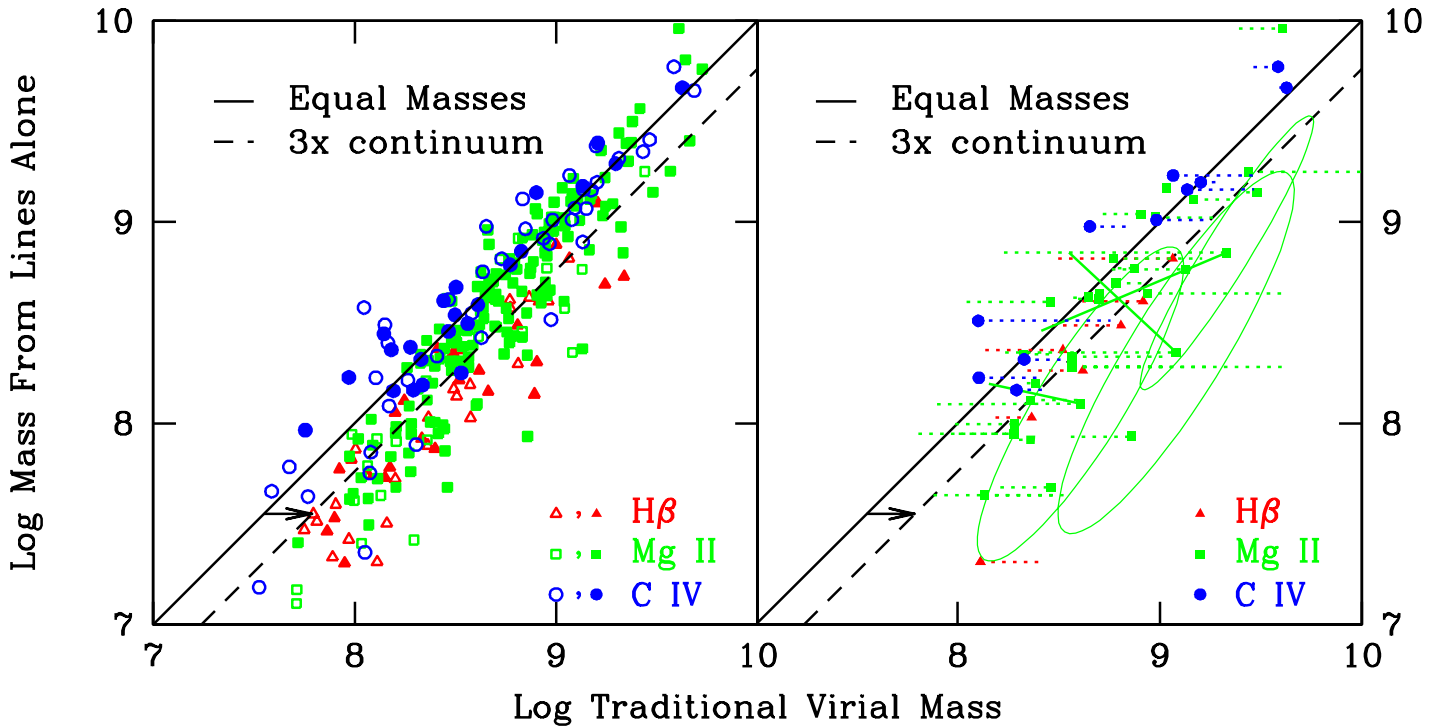}
\vspace*{-250pt}
\caption{On the left, masses from the lines are plotted against traditional virial masses for 
{\it Fermi} FSRQ. Most objects lie to the right of the 1:1 (solid) line,
indicating larger traditional (continuum) masses.  The dashed line shows the 
effect of a $3\times$ {\it NTD}. Closed symbols indicate objects with error $< 0.4$\,dex 
in both masses. In the right panel, objects with multiple observations are plotted. For 
the three objects with significant ($>2\sigma$) mass offset, we connect the two masses 
with a solid line, and plot the error ellipse around the primary observation. For others, 
we measure the continuum luminosity in both spectra, and mark the 
mass offset from continuum change by a dotted line.} 
\label{fig:masscomp}
\end{figure*}

\subsection{ Non-thermal Continuum Pollution}
\label{sec:nontherm}

In Figure \ref{fig:nontherm}, we report on the $NTD$ of 
our sample. An optically selected population, such as that of \citet{she11}, 
peaks at $\log NTD=0$, the line-predicted thermal flux is comparable to the measured
flux. Unsurprisingly the {\it Fermi} blazars extend to much larger $NTD$ levels.
This plays an important role in continuum-calibrated estimates of hole mass in \S \ref{sec:masses}.

We further find a weak, but interesting correlation between {\it NTD} and 
{\it Fermi} $\gamma$-ray spectral index. The {\it Fermi} FSRQ population has 
$<\gamma> = 2.44$, whereas the BL Lac population has $<\gamma>=2.04$. Thus
the $\gamma$-ray hardness is a good predictor of the continuum strength
in the broad band SED. Indeed, we find that the objects in our sample with the
largest {\it NTD} show relatively hard {\it Fermi} spectra. It will
be of interest to trace this trend deeper into the BL Lac population,
although {\it NTD} estimates will be more difficult for this sample. Note that
while actual BL Lacs (circles), unsurprisingly, are mostly in the high 
{\it NTD} regime, a number of nominal FSRQ are found in this region, as well.

\section{The Central Engine}
\label{sec:masses}

\subsection{Traditional Virial Mass Estimates}
\label{sec:traditional}

As in \citet{she11}, we estimate traditional virial black hole masses from a 
relation of the form:
\begin{equation}
\label{eqn:masstemplate}
\log\left(\frac{M_{BH}}{M_\odot}\right) =   a + b \log\left(\lambda L_\lambda\right) + 2 \log \left(\textrm{FWHM}\right)
\end{equation}
where $a$ and $b$ are calibrated from reverberation mapping for each line 
species \citep{mcl04}. The broad line FWHM, in km s$^{-1}$, and the continuum 
luminosity $\lambda L_\lambda$, in units of $10^{44}$ erg s$^{-1}$, are measured as 
described in \S \ref{sec:hb}, \S \ref{sec:MgII}, and \S \ref{sec:civ}.
Values of $a$ and $b$ are tabulated in Table \ref{tab:mass} from \citet{ves06} 
(hereafter VP06) for H $\beta$ and \ion{C}{4} and \citet{ves09} (hereafter, VO09) 
for \ion{Mg}{2}.
A scatter of $\sim 0.4$ dex has been inferred for virial masses in optically 
selected samples \citep{she11}.

For our {\it Fermi} sample we estimate mass from H $\beta$ for 50 objects, 
from \ion{Mg}{2} for 176 objects, 
and from \ion{C}{4} for 68 objects. For 39 and 50 objects respectively, we 
measure mass from both H $\beta$ and \ion{Mg}{2}, and from \ion{Mg}{2} and 
\ion{C}{4}. Masses are reported in Table \ref{tab:lowz} and Table \ref{tab:highz}.

We urge caution in a naive interpretation of these masses, as applied to blazers. Due to the significant $NTD$ of our sample, we expect continuum luminosity to not scale as in the reverberation mapping sample, as discussed below in \S \ref{sec:linesalone}. Further, preferential alignment and a non-spherical BLR yields systematically narrower lines, as discussed in \S \ref{sec:orientation}.

\subsection{Mass Estimates from Lines Alone}
\label{sec:linesalone}

\begin{deluxetable*}{lccccccc}
\tabletypesize{\small}
\tablecaption{Coefficients for Black Hole Mass Estimates }
\tablehead{
\colhead{Source} & \multicolumn{2}{c}{H $\beta$} & \multicolumn{2}{c}{\ion{Mg}{2}} & \multicolumn{2}{c}{\ion{C}{4}} & \colhead{Reference}\\
\colhead{} & \colhead{a} & \colhead{b} & \colhead{a} & \colhead{b} & \colhead{a} & \colhead{b} & \colhead{}
}
\startdata
Mass from Continuum & 0.672 & 0.61 & 0.505 & 0.62 & 0.660 & 0.53 & MD04 (H$\beta$,\ion{C}{4}), VO09 (\ion{Mg}{2})\\  
Mass from Lines  & $1.63 \pm .04$  & $0.49 \pm .03$ & $1.70 \pm .07$ & $ .63 \pm .00$ & $1.52 \pm .22$ & $.46 \pm .01$  & this work (H$\beta$), \citet{she11} (\ion{Mg}{2}, \ion{C}{4})\\  
\enddata
\tablecomments{Coefficients for the virial fitting relation in Equation \ref{eqn:masstemplate}. Coefficients derived from this work, \citet[MD04]{mcl04}, \citet[VO09]{ves09}, and \citet{she11}}
\label{tab:mass}
\end{deluxetable*}

\begin{figure}
\vspace{-15pt}
\hspace*{-20pt}
\epsscale{1.3}
\plotone{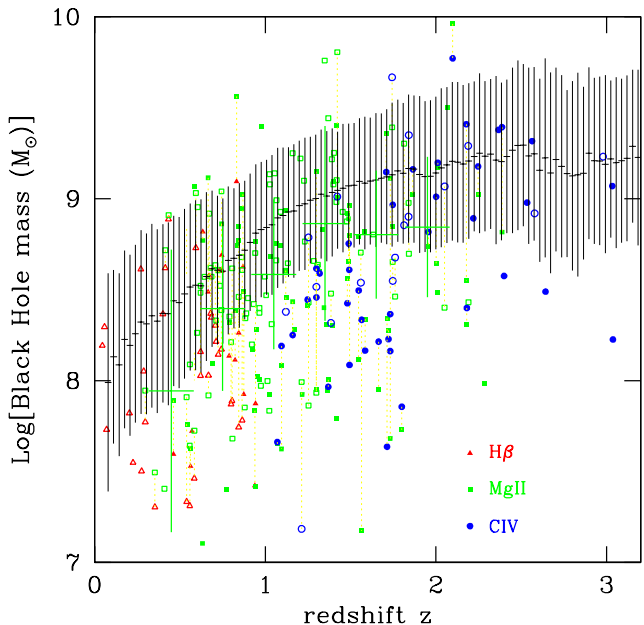}
\vspace{-25pt}
\caption{Blazar black hole (line) masses as a function of redshift.  Masses are 
over-plotted on the Sloan average quasar masses. Objects with $i < 19.1$ 
are plotted as open points. Large green crosses show the average MgII-estimated mass and range for our magnitude-cut Fermi blazar sample in six redshift bins.} 
\label{fig:mass}
\end{figure}

Bearing in mind the substantial {\it NTD} of much of our {\it Fermi} sample,
we wish to make mass estimates from the line luminosity and line kinematic width 
alone.  \citet{gre05} and \citet{kon06} calibrate such estimators from 
reverberation mapping samples. \citet{she11} calibrates estimators for 
\ion{Mg}{2} and \ion{C}{4} from the SDSS sample. We have augmented this, using 
the SDSS mass estimators and developed our own coefficients for H$\beta$ with
values consistent with the predicted continuum luminosity and mass estimates 
of \S \ref{sec:traditional}. The coefficients for these estimators are listed 
in Table \ref{tab:mass} for a formula of the form of Equation \ref{eqn:masstemplate}, 
replacing $\lambda L_\lambda$ with line luminosity in units of $10^{44}$ erg s$^{-1}$.

In Figure \ref{fig:masscomp} we compare the line-estimated masses for our 
radio-$\gamma$ selected sample with estimates from the traditional continuum
virial mass equations. We find the line masses are smaller than the traditional 
virial masses by 0.14 dex on average. This is a small, but significant effect.
More interesting is the comparison for different species and different epochs.
For example the nearby, low mass, low luminosity H$\beta$ mass sample shows 
a larger average line-mass decrease than the powerful \ion{C}{4} FSRQ. We attribute
this primarily to host light pollution in the former and thermal disk domination in the latter.

For those objects where we have multiple epochs of observation (see 
\S \ref{sec:mult} for details), we calculate the mass in both spectra. Due the 
statistical uncertainty in measuring the line kinematic width, and the large role 
that plays in mass calculations, in all but three cases, the measurement in 
the multiple spectra are not significantly different from that in the primary spectrum. 
In the right panel of Figure \ref{fig:masscomp}, we also plot estimated mass 
changes from continuum luminosity fluctuations in spectra too poorly measured 
to estimate line properties. The general trend is for variations to extend above the
line-determined values, which show smaller fluctuation (i.e. continuum variations
dominate and bias mass estimates, as expected).

\subsection{Comparison to Optically Selected Quasars}
\label{sec:masscomp}

Using the less biased masses derived using line strengths, we compare in Figure \ref{fig:mass} our 
radio-$\gamma$ sample to the optically selected SDSS QSO \citep{she11}. 
Interestingly, the mean mass varies with redshift in parallel to the QSO
sample, but lower. For the full {\it Fermi} sample the offset is 0.44 dex.
For the magnitude-cut sub-sample (directly comparable to the SDSS QSO), the
offset is 0.34 dex.

	The origin of this offset is unclear. Certainly the {\it NTD} of our
sample causes some intrinsically fainter, less massive blazars to be boosted into
our flux cut sample. Further, our {\it Fermi} objects sample a preferred 
orientation (see \S \ref{sec:orientation}).  Another possibility is that these 
sources are on average in a more active accretion state than the typical 
QSO (see \S \ref{sec:edd}). We also allow that the brighter blazars with with historical spectra in the literature may have systematically larger hole masses, decreasing the magnitude of this offset.

\subsection{Eddington Ratio}
\label{sec:edd}

\begin{figure}
\vspace{-15pt}
\hspace*{-5pt}
\epsscale{1.25}
\plotone{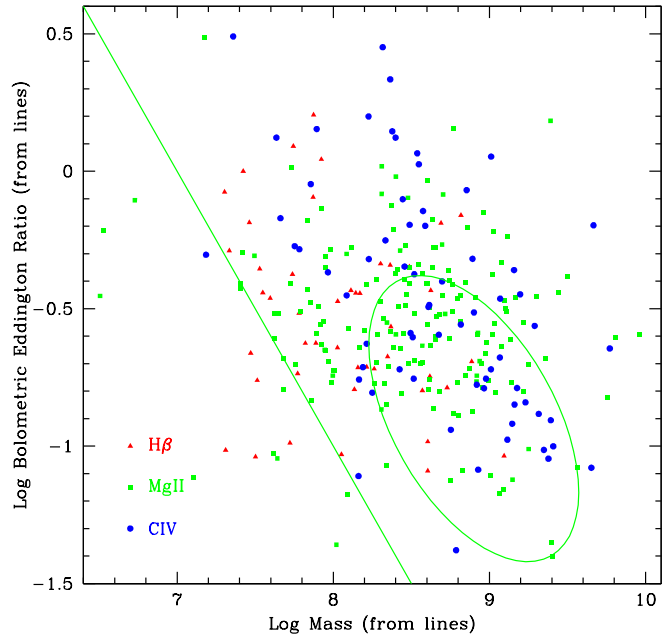}
\vspace{-15pt}
\caption{The Eddington ratio plotted against black hole mass. The green ellipse shows
the centroid and approximate 1$\sigma$ range for values determined
from \ion{Mg}{2} lines for the SDSS QSO sample. The green line shows the approximate 
flux limit for our sample.}
\label{fig:edd}
\end{figure}

Following the unified model \citep{urr95}, we expect the quasi-isotropic thermal
emissions of blazars to be similar to those in the underlying QSO population. 

To probe this, we measure the disc luminosity in Eddington units, $L_{Edd}$, using
line measurements alone. We first use the line fluxes to estimate local continuum
luminosities, as above, then convert these to bolometric fluxes using factors
from \citet{ric06} ($L_{bol}/L_{5100}=9.26$, $L_{bol}/L_{3000}=5.15$,
$L_{bol}/L_{1350} = 3.81$). This provides a bolometric thermal (= accretion
power) luminosity, independent of any non-thermal contribution. This luminosity
may be compared with the line-estimated hole masses to derive the Eddington ratio 
$L_{bol} / L_{Edd}$ for the population. Figure \ref{fig:edd} displays
this ratio against line-determined hole mass. Our sample has a systematically 
higher Eddington ratio and lower mass than the optically selected SDSS quasars. 
The radio-$\gamma$ selection probes an intrinsically more active population 
than an optically selected sample, both in flux limited samples.

\citet{ghi11} has proposed that FSRQ are more active accretors than BL Lacs. While
our high Eddington ratios for this FSRQ support this picture, we do not find any 
inverse correlation between Eddington ratio and non-thermal dominance as might be
expected if high values indicate a transition to a BL Lac-type state.

\section{Orientation of the BLR}
\label{sec:orientation}

The virial black hole mass estimates require that the broad line region 
(BLR) has an isotropic velocity distribution \citep{sal07}. In practice, there 
is a lack of consensus as to the shape of the BLR \citep{mcl04}. With a more 
disc-like BLR, objects observed perpendicular to the disc will have lower kinematic
FWHMs, and thus, lower inferred BH masses.  Following \citet{dec11}, we can interpret 
the FWHM offset of \S\ref{sec:sdsscomp} as a geometric effect and estimate the 
$f$-value, as a measure of the shape of the broad line region.

If we assume that the underlying broad line shape distribution is the same as that
of the SDSS QSOs, and use our $i < 19.1$ flux-cut sample, we find that our 
\ion{Mg}{2} and \ion{C}{4} lines are 0.14 dex narrower. Recalling that our line
widths seem to be systematically slightly lower than the SDSS line estimates by 0.04 dex, the true offset may be as high as 0.18 dex.

We assume we are probing a population with the same mass distribution as SDSS. Then, for a 0.14 dex offset our f-value is 1.38 $f_\textrm{SDSS}$, which is inconsistent with 
both a spherical BLR and a geometrically thin disk. In fact $f$ can be related to
the typical disk thickness ratio  H/R:
\begin{equation}
< f > = \left< 0.5 \left[\left(\frac{H}{R}\right)^2 + \sin^2\theta\right]^{-1/2} \right>
\end{equation}
where the average is taken over the observer inclination angle $\theta$. Since there is
good evidence that the {\it Fermi} FSRQs are highly aligned (within 5$^{\circ}$; Ajello et al, 
submitted to ApJ) by assuming that the SDSS QSO are viewed uniformly (in $\sin \theta$)
away from the obscuring equatorial torus (at $\theta > 60^\circ$), we obtain an estimate
for $H/R \sim 0.4$. Since the function is very flat at low $\theta$, the result is 
insensitive to the precise degree of {\it Fermi} FSRQs alignment. In fact the required
axis ratio only increases to 0.44 if the {\it Fermi} FSRQ are aligned within $2^\circ$. 
The decreased line width, and hence the decreased inferred hole mass
are fully consistent with the preferred alignment of these blazars as expected in the
Standard Model. Thus while we do not exclude underlying masses differences, our measurements
are best explained by similar optically selected and radio-$\gamma$ selected black hole mass
distributions, but with viewing angles strongly aligned to the disk (and jet) axis
for the latter.  As noted in \S \ref{sec:edd}, we do however find that
the FSRQ show a higher Eddington ratio than the typical QSO, so some difference
in accretion activity is indicated.

\section{Comparison with Radio Activity}
\label{sec:multiwavelength}

Since the {\it Fermi} FSRQ associations are largely selected from a sample of radio
bright, flat-spectrum sources, it may be useful to compare the optical spectroscopic
properties with radio measurements indicating strong non-thermal activity. One good
measure of this activity is the radio variability.  Much of the Dec $>$ -20 blazar sample 
has been monitored at 15\,GHz at OVRO since before the launch of {\it Fermi}. 
These data have been used to measure the intrinsic modulation index, a measure of
the activity:
\begin{equation}
\label{eqn:im}
\overline{m} = \frac{\sigma_0}{S_0}
\end{equation}
where $S_0$ and $\sigma_0$ are the intrinsic mean flux density and its standard deviation of the radio light curve, estimated from the data using a maximum-likelihood method \citep{ric11}. It has, for example, been found that this index is smaller for high redshift FSRQ than lower redshift BL Lacs.

\begin{figure}
\vspace{-15pt}
\hspace*{-20pt}
\epsscale{1.3}
\plotone{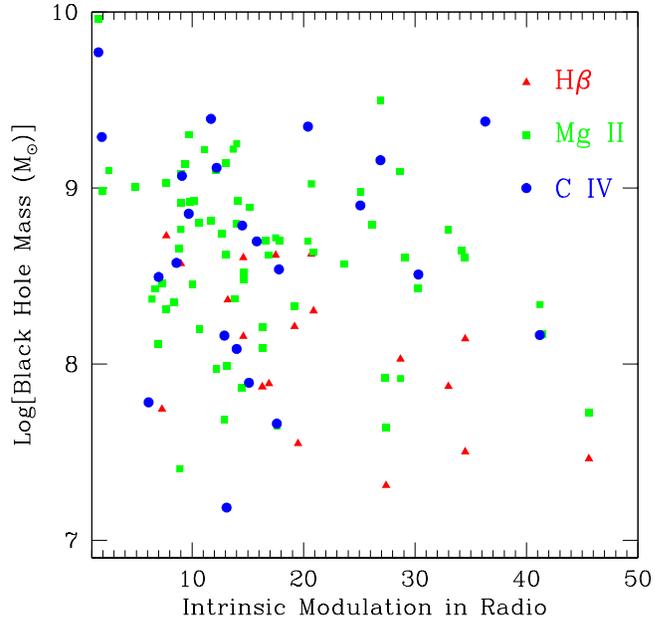}
\vspace{-30pt}
\caption{The black hole mass is uncorrelated to the intrinsic modulation in radio.} 
\label{fig:radio1}
\end{figure}

\begin{figure}
\vspace{-15pt}
\hspace*{-5pt}
\epsscale{1.2}
\plotone{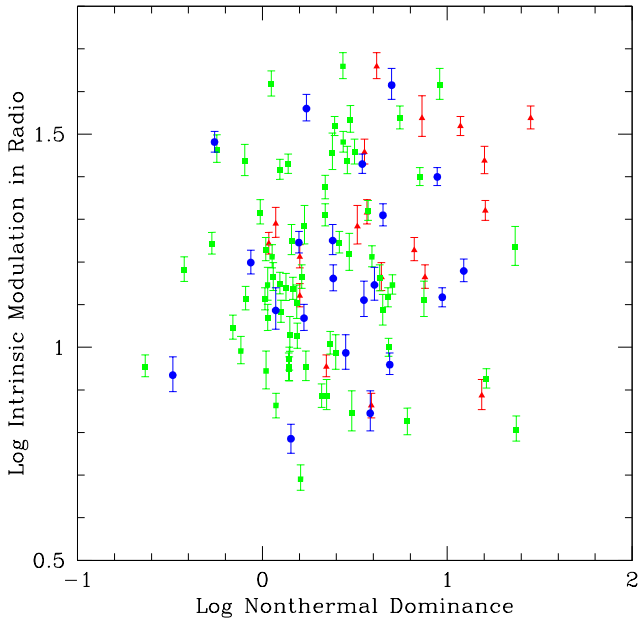}
\vspace{-10pt}
\caption{The intrinsic modulation is uncorrelated to the non-thermal dominance of the optical spectrum.} 
\label{fig:radio2}
\end{figure}

	We might, therefore, expect this modulation to correlate with either the black hole
mass Figure \ref{fig:radio1} or the {\it NTD} Figure \ref{fig:radio2}. The lowest intrinsic modulations
do appear among the highest mass holes, but there does not seem to be any increase
in modulation for FSRQ with the higher {\it NTD}. Thus jet variability does not seem
directly correlated with black hole and broad-line properties. It is possible that more
significant differences may be traced to the radio variability timescale, but this
requires longer radio monitoring and a more complete analysis of the flaring activity.

\section{Conclusions}

	We have used the optical spectral properties of our measured {\it Fermi} FSRQ sample
to characterize the state of the central engine. We find that the optical continuum is 
significantly augmented by non-thermal (synchrotron) emission, presumably associated with
the jet. In fact, such emission dominates the optical continuum for a third of our measured
FSRQ. Unsurprisingly the degree of $NTD$ correlates with the $\gamma$-ray
spectral index and the likelihood of BL Lac classification.

	To avoid bias from the additional continuum, we have developed estimates of the 
black hole mass and the Eddington luminosity ratio which only use the line properties.
These measurements show that the {\it Fermi} FSRQ are more active accretors than the 
bulk QSO population (at a given mass). We also find evidence for a significantly lower mean measured BH mass, even for a sample with a similar flux limit. Intriguingly, we find
that the redshift evolution of the mass tracks that of the full QSO population. Thus,
although non-thermal flux can pull less luminous, lower mass AGN into the sample, we
speculate that the primary effect is due to the high expected degree of alignment for
the FSRQ sample. If alignment is indeed the explanation, the data suggest a modest degree
(H/R $\sim 0.4$) of flattening in the broad line region, and that we probe a similar 
mean mass population (and evolution) as the bulk QSO distribution.

	While the present sample shows no strong correlation with the bulk radio variability,
we expect that a broader look at the {\it Fermi} blazar sample, including comparison
with the BL Lac properties and study of the BL Lac state-duty cycle and timescales 
for radio and optical modulation, will reveal additional correlations with the properties
of the central black hole and its surrounding broad line region. This extension to
the BL Lac portion of the sample is under way.

\acknowledgements

The Hobby*Eberly Telescope (HET) is a joint project of the University of Texas at
Austin, the Pennsylvania State University, Stanford University, Ludwig-Maximilians-Universitaet
Muenchen, and Georg-August-Universitaet Goettingen. The HET is named in honor of its principal
benefactors, William P. Hobby and Robert E. Eberly.
The Marcario Low Resolution Spectrograph is named for Mike Marcario of High Lonesome
Optics, who fabricated several optics for the instrument but died before its completion.
The LRS is a joint project of the Hobby*Eberly Telescope partnership and the Instituto de
Astronomıa de la Universidad Nacional Autonoma de Mexico.

Based on observations made with ESO Telescopes at the La Silla Paranal Observatory under programme 
077.B-0056
078.B-0275
079.B-0831
083.B-0460
084.B-0711
087.A-0573. GC acknowledges support from STFC grant ST/H002456/1

	We acknowledge support from NASA grants NNX09AW30G, NXX10AU09G and NAS5-00147. 
A.C.S.R. is also supported under grant AST-0808050.

{\it Facilities:} \facility{Fermi}, \facility{Hale (DBSP)}, \facility{HET}, \facility{KECK:I (LRIS)}, \facility{NTT}, \facility{VLT:Antu (FORS2)}.

\bibliography{references}

\begin{thebibliography}{32}
\expandafter\ifx\csname natexlab\endcsname\relax\def\natexlab#1{#1}\fi

\bibitem[{{Abdo} {et~al.}(2010{\natexlab{a}}){Abdo}, {Ackermann}, {Ajello},
  {Allafort}, {Antolini}, {Atwood}, {Axelsson}, {Baldini}, {Ballet},
  {Barbiellini}, \& et~al.}]{1FGL}
{Abdo}, A.~A., {et~al.} 2010{\natexlab{a}}, \apjs, 188, 405

\bibitem[{{Abdo} {et~al.}(2010{\natexlab{b}}){Abdo}, {Ackermann}, {Ajello},
  {Baldini}, {Ballet}, {Barbiellini}, {Bastieri}, {Bechtol}, {Bellazzini},
  {Berenji}, {Blandford}, {Bloom}, {Bonamente}, {Borgland}, {Bouvier},
  {Brandt}, {Bregeon}, {Brez}, {Brigida}, {Bruel}, {Buehler}, {Burnett},
  {Buson}, {Caliandro}, {Cameron}, {Cannon}, {Caraveo}, {Carrigan},
  {Casandjian}, {Cavazzuti}, {Cecchi}, {{\c C}elik}, {Celotti}, {Charles},
  {Chekhtman}, {Chen}, {Cheung}, {Chiang}, {Ciprini}, {Claus}, {Cohen-Tanugi},
  {Colafrancesco}, {Conrad}, {Davis}, {Dermer}, {de Angelis}, {de Palma},
  {Silva}, {Drell}, {Dubois}, {Favuzzi}, {Fegan}, {Ferrara}, {Fortin},
  {Frailis}, {Fukazawa}, {Fusco}, {Gargano}, {Gasparrini}, {Gehrels},
  {Germani}, {Giglietto}, {Giommi}, {Giordano}, {Giroletti}, {Glanzman},
  {Godfrey}, {Grandi}, {Grenier}, {Grove}, {Guillemot}, {Guiriec}, {Hadasch},
  {Hayashida}, {Hays}, {Horan}, {Hughes}, {Jackson}, {J{\'o}hannesson},
  {Johnson}, {Johnson}, {Kamae}, {Katagiri}, {Kataoka}, {Kn{\"o}dlseder},
  {Kuss}, {Lande}, {Latronico}, {Lee}, {Lemoine-Goumard}, {Llena Garde},
  {Longo}, {Loparco}, {Lott}, {Lovellette}, {Lubrano}, {Madejski}, {Makeev},
  {Malaguti}, {Mazziotta}, {McConville}, {McEnery}, {Michelson}, {Migliori},
  {Mitthumsiri}, {Mizuno}, {Monte}, {Monzani}, {Morselli}, {Moskalenko},
  {Murgia}, {Naumann-Godo}, {Nestoras}, {Nolan}, {Norris}, {Nuss}, {Ohsugi},
  {Okumura}, {Omodei}, {Orlando}, {Ormes}, {Paneque}, {Panetta}, {Parent},
  {Pelassa}, {Pepe}, {Persic}, {Pesce-Rollins}, {Piron}, {Porter}, {Rain{\`o}},
  {Rando}, {Razzano}, {Razzaque}, {Reimer}, {Reimer}, {Reyes}, {Roth},
  {Sadrozinski}, {Sanchez}, {Sander}, {Scargle}, {Sgr{\`o}}, {Siskind},
  {Smith}, {Spandre}, {Spinelli}, {Stawarz}, {Stecker}, {Strickman}, {Suson},
  {Takahashi}, {Tanaka}, {Thayer}, {Thayer}, {Thompson}, {Tibaldo}, {Torres},
  {Torresi}, {Tosti}, {Tramacere}, {Uchiyama}, {Usher}, {Vandenbroucke},
  {Vasileiou}, {Vilchez}, {Villata}, {Vitale}, {Waite}, {Wang}, {Winer},
  {Wood}, {Yang}, {Ylinen}, \& {Ziegler}}]{misaligned}
---. 2010{\natexlab{b}}, \apj, 720, 912

\bibitem[{{Abdo} {et~al.}(2010{\natexlab{c}}){Abdo}, {Ackermann}, {Ajello},
  {Allafort}, {Antolini}, {Atwood}, {Axelsson}, {Baldini}, {Ballet},
  {Barbiellini}, {Bastieri}, {Baughman}, {Bechtol}, {Bellazzini}, {Berenji},
  {Blandford}, {Bloom}, {Bogart}, {Bonamente}, {Borgland}, {Bouvier},
  {Bregeon}, {Brez}, {Brigida}, {Bruel}, {Buehler}, {Burnett}, {Buson},
  {Caliandro}, {Cameron}, {Cannon}, {Caraveo}, {Carrigan}, {Casandjian},
  {Cavazzuti}, {Cecchi}, {{\c C}elik}, {Celotti}, {Charles}, {Chekhtman},
  {Chen}, {Cheung}, {Chiang}, {Ciprini}, {Claus}, {Cohen-Tanugi}, {Conrad},
  {Costamante}, {Cotter}, {Cutini}, {D'Elia}, {Dermer}, {de Angelis}, {de
  Palma}, {De Rosa}, {Digel}, {Silva}, {Drell}, {Dubois}, {Dumora}, {Escande},
  {Farnier}, {Favuzzi}, {Fegan}, {Ferrara}, {Focke}, {Fortin}, {Frailis},
  {Fukazawa}, {Funk}, {Fusco}, {Gargano}, {Gasparrini}, {Gehrels}, {Germani},
  {Giebels}, {Giglietto}, {Giommi}, {Giordano}, {Giroletti}, {Glanzman},
  {Godfrey}, {Grandi}, {Grenier}, {Grondin}, {Grove}, {Guiriec}, {Hadasch},
  {Harding}, {Hayashida}, {Hays}, {Healey}, {Hill}, {Horan}, {Hughes},
  {Iafrate}, {Itoh}, {J{\'o}hannesson}, {Johnson}, {Johnson}, {Johnson},
  {Johnson}, {Kamae}, {Katagiri}, {Kataoka}, {Kawai}, {Kerr}, {Kn{\"o}dlseder},
  {Kuss}, {Lande}, {Latronico}, {Lavalley}, {Lemoine-Goumard}, {Llena Garde},
  {Longo}, {Loparco}, {Lott}, {Lovellette}, {Lubrano}, {Madejski}, {Makeev},
  {Malaguti}, {Massaro}, {Mazziotta}, {McConville}, {McEnery}, {McGlynn},
  {Michelson}, {Mitthumsiri}, {Mizuno}, {Moiseev}, {Monte}, {Monzani},
  {Morselli}, {Moskalenko}, {Murgia}, {Nolan}, {Norris}, {Nuss}, {Ohno},
  {Ohsugi}, {Omodei}, {Orlando}, {Ormes}, {Ozaki}, {Paneque}, {Panetta},
  {Parent}, {Pelassa}, {Pepe}, {Pesce-Rollins}, {Piranomonte}, {Piron},
  {Porter}, {Rain{\`o}}, {Rando}, {Razzano}, {Reimer}, {Reimer}, {Reposeur},
  {Ripken}, {Ritz}, {Rodriguez}, {Romani}, {Roth}, {Ryde}, {Sadrozinski},
  {Sanchez}, {Sander}, {Saz Parkinson}, {Scargle}, {Sgr{\`o}}, {Shaw},
  {Siskind}, {Smith}, {Spandre}, {Spinelli}, {Starck}, {Stawarz}, {Strickman},
  {Suson}, {Tajima}, {Takahashi}, {Takahashi}, {Tanaka}, {Taylor}, {Thayer},
  {Thayer}, {Thompson}, {Tibaldo}, {Torres}, {Tosti}, {Tramacere}, {Ubertini},
  {Uchiyama}, {Usher}, {Vasileiou}, {Vilchez}, {Villata}, {Vitale}, {Waite},
  {Wallace}, {Wang}, {Winer}, {Wood}, {Yang}, {Ylinen}, \& {Ziegler}}]{1LAC}
---. 2010{\natexlab{c}}, \apj, 715, 429

\bibitem[{{Appenzeller} {et~al.}(1998){Appenzeller}, {Fricke}, {F{\"u}rtig},
  {G{\"a}ssler}, {H{\"a}fner}, {Harke}, {Hess}, {Hummel}, {J{\"u}rgens},
  {Kudritzki}, {Mantel}, {Meisl}, {Muschielok}, {Nicklas}, {Rupprecht},
  {Seifert}, {Stahl}, {Szeifert}, \& {Tarantik}}]{app98}
{Appenzeller}, I., {et~al.} 1998, The Messenger, 94, 1

\bibitem[{{Atwood} {et~al.}(2009){Atwood}, {Abdo}, {Ackermann}, {Althouse},
  {Anderson}, {Axelsson}, {Baldini}, {Ballet}, {Band}, {Barbiellini}, \&
  et~al.}]{atw09}
{Atwood}, W.~B., {et~al.} 2009, \apj, 697, 1071

\bibitem[{{Bohlin}(2007)}]{boh07}
{Bohlin}, R.~C. 2007, in Astronomical Society of the Pacific Conference Series,
  Vol. 364, The Future of Photometric, Spectrophotometric and Polarimetric
  Standardization, ed. {C.~Sterken}, 315--

\bibitem[{{Buzzoni} {et~al.}(1984){Buzzoni}, {Delabre}, {Dekker}, {Dodorico},
  {Enard}, {Focardi}, {Gustafsson}, {Nees}, {Paureau}, \& {Reiss}}]{buz84}
{Buzzoni}, B., {et~al.} 1984, The Messenger, 38, 9

\bibitem[{{Decarli} {et~al.}(2011){Decarli}, {Dotti}, \& {Treves}}]{dec11}
{Decarli}, R., {Dotti}, M., \& {Treves}, A. 2011, \mnras, 413, 39

\bibitem[{{Dekker} {et~al.}(1986){Dekker}, {Delabre}, \& {Dodorico}}]{dek86}
{Dekker}, H., {Delabre}, B., \& {Dodorico}, S. 1986, in Presented at the
  Society of Photo-Optical Instrumentation Engineers (SPIE) Conference, Vol.
  627, Society of Photo-Optical Instrumentation Engineers (SPIE) Conference
  Series, ed. {D.~L.~Crawford}, 339--348

\bibitem[{{Ghisellini} {et~al.}(2011){Ghisellini}, {Tavecchio}, {Foschini}, \&
  {Ghirlanda}}]{ghi11}
{Ghisellini}, G., {Tavecchio}, F., {Foschini}, L., \& {Ghirlanda}, G. 2011,
  \mnras, 414, 2674

\bibitem[{{Greene} \& {Ho}(2005)}]{gre05}
{Greene}, J.~E., \& {Ho}, L.~C. 2005, \apj, 630, 122

\bibitem[{{Healey} {et~al.}(2008){Healey}, {Romani}, {Cotter}, {Michelson},
  {Schlafly}, {Readhead}, {Giommi}, {Chaty}, {Grenier}, \&
  {Weintraub}}]{cgrabs}
{Healey}, S.~E., {et~al.} 2008, \apjs, 175, 97

\bibitem[{Jones {et~al.}(2001--)Jones, Oliphant, Peterson, {et~al.}}]{scipy}
Jones, E., Oliphant, T., Peterson, P., {et~al.} 2001--, {SciPy}: Open source
  scientific tools for {Python}

\bibitem[{{Kong} {et~al.}(2006){Kong}, {Wu}, {Wang}, \& {Han}}]{kon06}
{Kong}, M.-Z., {Wu}, X.-B., {Wang}, R., \& {Han}, J.-L. 2006, cjaa, 6, 396

\bibitem[{{Krisciunas} {et~al.}(1987){Krisciunas}, {Sinton}, {Tholen},
  {Tokunaga}, {Golisch}, {Griep}, {Kaminski}, {Impey}, \& {Christian}}]{kri87}
{Krisciunas}, K., {et~al.} 1987, \pasp, 99, 887

\bibitem[{{Kurtz} \& {Mink}(1998)}]{rvsao}
{Kurtz}, M.~J., \& {Mink}, D.~J. 1998, \pasp, 110, 934

\bibitem[{{McLure} \& {Dunlop}(2004)}]{mcl04}
{McLure}, R.~J., \& {Dunlop}, J.~S. 2004, \mnras, 352, 1390

\bibitem[{{Oke}(1990)}]{oke90}
{Oke}, J.~B. 1990, \aj, 99, 1621

\bibitem[{{Richards} {et~al.}(2006){Richards}, {Lacy}, {Storrie-Lombardi},
  {Hall}, {Gallagher}, {Hines}, {Fan}, {Papovich}, {Vanden Berk}, {Trammell},
  {Schneider}, {Vestergaard}, {York}, {Jester}, {Anderson}, {Budav{\'a}ri}, \&
  {Szalay}}]{ric06}
{Richards}, G.~T., {et~al.} 2006, \apjs, 166, 470

\bibitem[{{Richards} {et~al.}(2011){Richards}, {Max-Moerbeck}, {Pavlidou},
  {King}, {Pearson}, {Readhead}, {Reeves}, {Shepherd}, {Stevenson},
  {Weintraub}, {Fuhrmann}, {Angelakis}, {Zensus}, {Healey}, {Romani}, {Shaw},
  {Grainge}, {Birkinshaw}, {Lancaster}, {Worrall}, {Taylor}, {Cotter}, \&
  {Bustos}}]{ric11}
{Richards}, J.~L., {et~al.} 2011, \apjs, 194, 29

\bibitem[{{Salviander} {et~al.}(2007){Salviander}, {Shields}, {Gebhardt}, \&
  {Bonning}}]{sal07}
{Salviander}, S., {Shields}, G.~A., {Gebhardt}, K., \& {Bonning}, E.~W. 2007,
  \apj, 662, 131

\bibitem[{{Schlegel} {et~al.}(1998){Schlegel}, {Finkbeiner}, \&
  {Davis}}]{sch98}
{Schlegel}, D.~J., {Finkbeiner}, D.~P., \& {Davis}, M. 1998, \apj, 500, 525

\bibitem[{{Schneider} {et~al.}(2010){Schneider}, {Richards}, {Hall}, {Strauss},
  {Anderson}, {Boroson}, {Ross}, {Shen}, {Brandt}, {Fan}, {Inada}, {Jester},
  {Knapp}, {Krawczyk}, {Thakar}, {Vanden Berk}, {Voges}, {Yanny}, {York},
  {Bahcall}, {Bizyaev}, {Blanton}, {Brewington}, {Brinkmann}, {Eisenstein},
  {Frieman}, {Fukugita}, {Gray}, {Gunn}, {Hibon}, {Ivezi{\'c}}, {Kent}, {Kron},
  {Lee}, {Lupton}, {Malanushenko}, {Malanushenko}, {Oravetz}, {Pan}, {Pier},
  {Price}, {Saxe}, {Schlegel}, {Simmons}, {Snedden}, {SubbaRao}, {Szalay}, \&
  {Weinberg}}]{sch10}
{Schneider}, D.~P., {et~al.} 2010, \aj, 139, 2360

\bibitem[{{Shen} {et~al.}(2011){Shen}, {Richards}, {Strauss}, {Hall},
  {Schneider}, {Snedden}, {Bizyaev}, {Brewington}, {Malanushenko},
  {Malanushenko}, {Oravetz}, {Pan}, \& {Simmons}}]{she11}
{Shen}, Y., {et~al.} 2011, \apjs, 194, 45

\bibitem[{{Szokoly} {et~al.}(2004){Szokoly}, {Bergeron}, {Hasinger}, {Lehmann},
  {Kewley}, {Mainieri}, {Nonino}, {Rosati}, {Giacconi}, {Gilli}, {Gilmozzi},
  {Norman}, {Romaniello}, {Schreier}, {Tozzi}, {Wang}, {Zheng}, \&
  {Zirm}}]{fors}
{Szokoly}, G.~P., {et~al.} 2004, \apjs, 155, 271

\bibitem[{{Tody}(1986)}]{tod86}
{Tody}, D. 1986, in Society of Photo-Optical Instrumentation Engineers (SPIE)
  Conference Series, Vol. 627, Society of Photo-Optical Instrumentation
  Engineers (SPIE) Conference Series, ed. {D.~L.~Crawford}, 733--

\bibitem[{{Urry} \& {Padovani}(1995)}]{urr95}
{Urry}, C.~M., \& {Padovani}, P. 1995, \pasp, 107, 803

\bibitem[{{Valdes}(1986)}]{val86}
{Valdes}, F. 1986, in Presented at the Society of Photo-Optical Instrumentation
  Engineers (SPIE) Conference, Vol. 627, Society of Photo-Optical
  Instrumentation Engineers (SPIE) Conference Series, ed. {D.~L.~Crawford},
  749--756

\bibitem[{{Valdes}(1992)}]{val92}
{Valdes}, F. 1992, in Astronomical Society of the Pacific Conference Series,
  Vol.~25, Astronomical Data Analysis Software and Systems I, ed.
  {D.~M.~Worrall, C.~Biemesderfer, \& J.~Barnes}, 417--

\bibitem[{{Vestergaard} \& {Osmer}(2009)}]{ves09}
{Vestergaard}, M., \& {Osmer}, P.~S. 2009, \apj, 699, 800

\bibitem[{{Vestergaard} \& {Peterson}(2006)}]{ves06}
{Vestergaard}, M., \& {Peterson}, B.~M. 2006, \apj, 641, 689

\bibitem[{{Vestergaard} \& {Wilkes}(2001)}]{ves01}
{Vestergaard}, M., \& {Wilkes}, B.~J. 2001, \apjs, 134, 1

\end{thebibliography}

\clearpage

\LongTables

\begin{landscape}

\begin{deluxetable}{lllllllllllll}
\tablecolumns{13}
\tabletypesize{\tiny}
\tablecaption{Line Properties for $z < 1$ Broad Line 1LAC AGN}
\tablehead{
\colhead{} & \colhead{} & \colhead{} &\colhead{} & \multicolumn{4}{c}{H$\beta$} & \multicolumn{4}{c}{\ion{Mg}{2}} & \colhead{}  \\
\cline{5-8} \cline{9-12}
\colhead{Name} &  \colhead{z} & \colhead{$F_{\nu,10^{14.7}}$} & \colhead{$\alpha$} & \colhead{$L_{5100}$} &  \colhead{$L_{H \beta}$} & \colhead{FWHM} & \colhead{Mass} & \colhead{$L_{3000}$} & \colhead{$L_{\textrm{\ion{Mg}{2}}}$}& \colhead{FWHM} & \colhead{Mass} & \colhead{Telescope} \\
\colhead{} & \multicolumn{3}{c}{$10^{-28}$erg cm$^{-2}$s$^{-1}$Hz$^{-1}$}  & \colhead{erg s$^{-1}$} &  \colhead{erg s$^{-1}$} & \colhead{kms$^{-1}$} & \colhead{$\log \frac{M}{M_\odot}$} &  \colhead{erg s$^{-1}$} &  \colhead{erg s$^{-1}$} & \colhead{kms$^{-1}$}  & \colhead{$\log \frac{M}{M_\odot}$} & \colhead{}
}\startdata
J0004$-$4736 & 0.880 & 11.4$\pm$0.0 & $-$0.99$\pm$0.08 & ... & ... & ... & ... & 45.339$\pm$0.011 & 42.896$\pm$0.151 & 2700$\pm$500 & 7.85$\pm$0.36 & NTT\\
J0008+1450 & 0.045 & 73.8$\pm$0.1 & $-$1.79$\pm$0.23 & 43.216$\pm$0.002 & 40.921$\pm$0.120 & 10800$\pm$2600 & 8.19$\pm$0.48 & ... & ... & ... & ... & SDSS\\
J0017$-$0512 & 0.226 & 30.4$\pm$0.1 & $-$0.46$\pm$0.36 & 44.353$\pm$0.004 & 42.388$\pm$0.107 & 2300$\pm$500 & 7.55$\pm$0.45 & ... & ... & ... & ... & HET\\
J0024+0349 & 0.545 & 27.1$\pm$0.0 & $-$1.88$\pm$0.08 & ... & ... & ... & ... & 45.110$\pm$0.001 & 42.585$\pm$0.057 & 3000$\pm$200 & 7.76$\pm$0.14 & WMKO\\
J0043+3426 & 0.966 & 6.6$\pm$0.0 & $-$1.24$\pm$0.03 & ... & ... & ... & ... & 45.201$\pm$0.006 & 42.807$\pm$0.072 & 3400$\pm$300 & 8.01$\pm$0.16 & WMKO\\
J0050$-$0452 & 0.922 & 7.7$\pm$0.1 & $-$0.92$\pm$0.19 & ... & ... & ... & ... & 45.198$\pm$0.013 & 43.140$\pm$0.141 & 3300$\pm$600 & 8.20$\pm$0.34 & HET\\
J0102+4214 & 0.874 & 30.1$\pm$0.1 & $-$1.63$\pm$0.86 & 46.001$\pm$0.004 & 43.460$\pm$0.038 & 1900$\pm$200 & 7.92$\pm$0.16 & 45.409$\pm$0.007 & 43.593$\pm$0.072 & 3300$\pm$300 & 8.49$\pm$0.17 & WMKO\\
J0102+5824 & 0.644 & 167.0$\pm$0.3 & $-$1.25$\pm$0.36 & ... & ... & ... & ... & 46.093$\pm$0.076 & 43.449$\pm$0.256 & 4100$\pm$1200 & 8.57$\pm$0.61 & HET\\
J0112+2244 & 0.265 & 18.6$\pm$0.2 & $-$2.71$\pm$0.79 & ... & ... & ... & ... & ... & ... & ... & ... & HET\\
J0113+1324 & 0.685 & 15.2$\pm$0.0 & $-$0.01$\pm$0.26 & 45.063$\pm$0.006 & 43.094$\pm$0.062 & 3800$\pm$500 & 8.35$\pm$0.25 & 45.263$\pm$0.006 & 43.389$\pm$0.082 & 4300$\pm$400 & 8.57$\pm$0.19 & SDSS
\enddata
\label{tab:lowz}
\tablecomments{Table \ref{tab:lowz} is published in its entirety in the electronic edition of this journal; A portion is shown here for guidance regarding its form and content.}
\end{deluxetable}

\begin{deluxetable}{lllllllllllll}
\tablecolumns{13}
\tabletypesize{\tiny}
\tablecaption{Line Properties for $z > 1$ Broad Line 1LAC AGN}
\tablehead{
\colhead{} & \colhead{} & \colhead{} &\colhead{} & \multicolumn{4}{c}{\ion{Mg}{2}} & \multicolumn{4}{c}{\ion{C}{4}} & \colhead{}  \\
\cline{5-8} \cline{9-12}
\colhead{Name} &  \colhead{z} & \colhead{$F_{\nu,10^{14.7}}$} & \colhead{$\alpha$} & \colhead{$L_{5100}$} &  \colhead{$L_{\textrm{\ion{Mg}{2}}}$} & \colhead{FWHM} & \colhead{Mass} & \colhead{$L_{3000}$} & \colhead{$L_{\textrm{\ion{C}{4}}}$}& \colhead{FWHM} & \colhead{Mass} & \colhead{Telescope} \\
\colhead{} & \multicolumn{3}{c}{$10^{-28}$erg cm$^{-2}$s$^{-1}$Hz$^{-1}$}  & \colhead{erg s$^{-1}$} &  \colhead{erg s$^{-1}$} & \colhead{kms$^{-1}$} & \colhead{$\log \frac{M}{M_\odot}$} &  \colhead{erg s$^{-1}$} &  \colhead{erg s$^{-1}$} & \colhead{kms$^{-1}$}  & \colhead{$\log \frac{M}{M_\odot}$} & \colhead{}
}\startdata
J0011+0057 & 1.493 & 5.2$\pm$0.1 & $-$0.51$\pm$0.05 & 45.486$\pm$0.021 & 43.541$\pm$0.081 & 4900$\pm$500 & 8.80$\pm$0.19 & 45.800$\pm$0.069 & 43.493$\pm$0.435 & 2500$\pm$1300 & 8.09$\pm$1.02 & SDSS\\
J0015+1700 & 1.709 & 8.0$\pm$0.0 & $-$2.75$\pm$0.38 & 46.070$\pm$0.004 & 44.178$\pm$0.043 & 5900$\pm$300 & 9.36$\pm$0.10 & 45.818$\pm$0.016 & 44.180$\pm$0.068 & 5900$\pm$500 & 9.15$\pm$0.17 & WMKO\\
J0023+4456 & 2.023 : & 1.3$\pm$0.1 & $-$0.20$\pm$0.50 & ... & ... & ... & ... & 45.213$\pm$1.000 & 43.336$\pm$0.875 & 1900$\pm$2200 & 7.78$\pm$2.29 & HET\\
J0042+2320 & 1.426 & 5.1$\pm$0.1 & $-$1.40$\pm$0.08 & 45.533$\pm$0.017 & 43.412$\pm$0.137 & 6900$\pm$1100 & 9.01$\pm$0.32 & ... & ... & ... & ... & HET\\
J0044$-$8422 & 1.032 & 9.7$\pm$0.0 & $-$0.14$\pm$0.03 & 45.401$\pm$0.007 & 43.667$\pm$0.113 & 3900$\pm$500 & 8.68$\pm$0.26 & ... & ... & ... & ... & VLT\\
J0048+2235 & 1.161 & 11.5$\pm$0.0 & $-$1.34$\pm$0.18 & 45.639$\pm$0.002 & 43.143$\pm$0.049 & 4300$\pm$200 & 8.43$\pm$0.11 & 45.916$\pm$0.027 & 43.272$\pm$0.115 & 3400$\pm$500 & 8.25$\pm$0.28 & WMKO\\
J0058+3311 & 1.369 & 3.1$\pm$0.1 & $-$2.49$\pm$0.37 & 45.331$\pm$0.003 & 42.805$\pm$0.054 & 3400$\pm$200 & 8.01$\pm$0.13 & 45.177$\pm$0.035 & 43.451$\pm$0.076 & 2200$\pm$200 & 7.97$\pm$0.19 & WMKO\\
J0104$-$2416 & 1.747 & 8.8$\pm$0.0 & $-$0.81$\pm$0.22 & 45.817$\pm$0.010 & 43.613$\pm$0.111 & 5000$\pm$600 & 8.85$\pm$0.26 & 45.989$\pm$0.025 & 44.121$\pm$0.292 & 4900$\pm$1700 & 8.97$\pm$0.69 & NTT\\
J0157$-$4614 & 2.287 & 1.9$\pm$0.0 & $-$1.07$\pm$0.33 & 45.632$\pm$0.030 & 43.216$\pm$0.171 & 2500$\pm$500 & 7.98$\pm$0.40 & 45.522$\pm$1.000 & 44.083$\pm$7.015 & 3000$\pm$24900 & 8.52$\pm$20.15 & VLT\\
J0226+0937 & 2.605 & 16.9$\pm$0.0 & $-$0.58$\pm$0.26 & ... & ... & ... & ... & 46.811$\pm$0.006 & 44.582$\pm$0.720 & 8500$\pm$7500 & 9.65$\pm$1.76 & P200
\enddata
\label{tab:highz}
\tablecomments{Table \ref{tab:highz} is published in its entirety in the electronic edition of this journal; A portion is shown here for guidance regarding its form and content.}
\end{deluxetable}
\clearpage

\end{landscape}

\end{document}